\documentclass[twocolumn,superscriptaddress,showpacs,nofootinbib]{revtex4-1}

\usepackage[colorlinks,pdfstartview=FitH]{hyperref}
\hypersetup{linkcolor=blue,citecolor=blue,filecolor=black,urlcolor=blue}

\usepackage{amsmath,amssymb,bm}
\usepackage{slashed}
\usepackage{graphicx}
\usepackage{amsfonts}
\usepackage{color}

\usepackage{amsfonts}
\usepackage{bbm}
\usepackage{latexsym}

\begin{document}

\title{Tamed loops: A way to obtain finite loop results without UV divergences}

\author{Lian-Bao Jia}  \email{jialb@mail.nankai.edu.cn}
\affiliation{
School of Mathematics and Physics, Southwest University of Science and Technology, Mianyang 621010, China}
\affiliation{School of Physics, Nankai University, Tianjin 300071, China}
\affiliation{Department of Physics, Chongqing Key Laboratory for Strongly Coupled Physics, Chongqing University, Chongqing 401331, China}

\begin{abstract}

For loops with UV divergences, assuming that the physical contributions of loops from UV regions are insignificant, a method of UV-free scheme described by an equation is introduced to derive loop results without UV divergences in calculations, i.e., a route of the analytic continuation $\mathcal{T}_\mathrm{F} \to \mathcal{T}_\mathrm{P}$ besides the traditional route $\infty - \infty$ in mathematical structure. This scheme provides a new perspective to an open question of the hierarchy problem of Higgs mass, i.e., an alternative interpretation without fine-tuning within the standard model.

\end{abstract}


\maketitle

\section{Introduction}

In quantum field theory, Feynman diagrams are used to describe perturbative contributions to the transition amplitudes of particle interactions, including tree and loop diagrams. For a loop diagram, the four-momentum of particles in the loop is not uniquely determined by the conservation of energy and momentum, and there is a free momentum $k^\mu$ in the loop. All possibilities contribute equally, and the evaluation is often ultraviolet (UV) divergence when we directly integrate over all possible $k^\mu$ that could travel around the loop. Hence, infinities from loop integrals at large energy and momentum regions ($k^\mu \to \infty$) indicate that constructions of loop contributions need further improvement \cite{DiracTheEO}.

The actual physics is obscured by infinities. How to make sense of infinities and get physical quantities when evaluating loop integrations? The first step of a paradigm approach is to make UV divergences mathematically expressed through regularization, followed by canceling UV divergences by renormalization with counterterms introduced. In Pauli-Villars regularization \cite{Pauli:1949zm}, massive fictitious particles are involved to cancel out divergences at large momenta. A popular method is dimensional regularization \cite{tHooft:1972tcz}, and a fictitious fractional number of spacetime dimensions is introduced into the integral (see e.g. Refs. \cite{Hadamard1932,Wilson:1974sk,Callan:1970yg,Symanzik:1970rt,Freedman:1991tk,tHooft:2004bkn,Liu:2022mfb,Mooij:2021ojy} for more methods). In the scheme of regularization followed by renormalization, the actual physics is extracted from infinities via $\infty - \infty =$ finite physical results. With this method, for example, the electron anomalous magnetic moment predicted by the standard model (SM) \cite{Aoyama:2012wj,Laporta:2017okg,Aoyama:2017uqe,Volkov:2019phy} agrees with the value measured by experiments \cite{Parker:2018vye,Morel:2020dww,Fan:2022eto} at an accuracy of $10^{-12}$.

There are two types of UV divergences, i.e., logarithmic divergence and power-law divergence. Despite the success of the regularization and renormalization procedure for logarithmic divergences, the feeling remains that there ought to be an economic way to acquire loop contributions. If we believe physical contributions from loops are finite, then an open question is how to find an appropriate way to directly obtain physically finite results without UV divergences. This is of our concern in this paper. A new method is explored here to derive finite loop contributions without UV divergences, and applications of the method are discussed.

\section{New method for loops}

As described in the Introduction, UV divergences of loops indicate that the transition amplitudes directly obtained are not well-defined in these cases. For this issue, a presumption on loops is proposed, i.e. the physical contributions of loops are locally finite with contributions from UV regions being insignificant. Hence, we assume that the physical transition amplitude $\mathcal{T}_\mathrm{P}$ with propagators can be described by an equation of
\begin{eqnarray}
\mathcal{T}_\mathrm{P} \! = \! \Big [ \mathrm{\int} \mathrm{d}\xi_1 \cdots \mathrm{d}\xi_i \frac{\partial  \mathcal{T}_\mathrm{F} (\xi_1,\cdots,\xi_i )}{\partial \xi_1 \cdots \partial \xi_i} \Big]_{\{\xi_1, \cdots, \xi_i\} \to 0} \! + \! C \, ,     \label{new-pro}
\end{eqnarray}
where a Feynman-like amplitude $\mathcal{T}_\mathrm{F} (\xi_1,\cdots,\xi_i )$ is introduced, which is written by Feynman rules just with parameters $\xi_1,\cdots,\xi_i$ added into denominators of propagators. This is suitable for the $k$-dependent integrand in a loop. For the antiderivative over the parameter $\xi$, here we introduce the core part induced by $\xi$, i.e., the primary antiderivative $[ \mathrm{\int} \mathrm{d}\xi_1 \cdots \mathrm{d}\xi_i \frac{\partial \mathcal{T}_\mathrm{F} (\xi_1,\cdots,\xi_i )}{\partial \xi_1 \cdots \partial \xi_i}]$ (defined as the antiderivative composed solely of $\xi$ terms), and the constant term is absorbed into the boundary constant $C$ which is related to the transition process (for a toy example of $\mathcal{T}_\mathrm{F} = \int_0^\infty \!\!\frac{1}{x+a} \mathrm{d}x$ with $a >$0, one has $\mathcal{T}_\mathrm{P} =$ [$\int \!\mathrm{d}\xi \int_0^\infty \!\!\frac{-1}{(x+a+\xi)^2} \mathrm{d}x]_{\xi \to 0}$ +$C$ \!=\! $[\log \frac{1}{a+\xi}]_{\xi \to 0}$ \!+$C$, and the final result of the primary antiderivative is $\log \frac{1}{a}$). After integral, $\mathcal{T}_\mathrm{P}$ will be obtained in the limit of parameters $\xi_1 \to 0$, $\cdots$, $\xi_i \to 0$ (a general form of the limit is $\xi = \rho \mathrm{e}^{\mathrm{i} \theta}$ with $\rho \to 0$). If Eq. (\ref{new-pro}) is applied to tree-level and loop-level processes without UV divergences, or loops with the primary antiderivative being zero, $C =$ 0 is adopted. For loop processes with UV divergences, $C$ can be set by renormalization conditions (or think of it as physical normalization conditions), symmetries and naturalness, which can be considered as the reference value of the primary antiderivative at given reference energy points (in the above example, if the result being zero at $a$=2 is adopted, $C = \log 2$ is obtained and the final result is $\mathcal{T}_\mathrm{P} = \log \frac{2}{a}$). The number of the parameter $\xi_i$ introduced is as few as possible in the case of the loop integral becoming UV-converged. For a loop with UV divergences, one parameter $\xi$ is introduced for logarithmic divergence, and two $\xi$ parameters are introduced for quadratic divergence (three $\xi$ parameters needed at most for a loop being converged). For multi-loops, a set of $\xi$ parameters is introduced for loops being converged. The method above is the UV-free scheme.

The loop calculation involves the following steps: \\
(a) Write down the transition amplitude $\mathcal{T}_\mathrm{F}$ by the Feynman rules. \\ (b) Add $\xi$ to the denominator of a propagator in the loop and take the derivative a sufficient number of times; Feynman parameterization is required for cases involving multiple propagators. \\ (c) Integrate over the loop momentum to obtain a UV-convergent result. \\ (d) Evaluate the antiderivative with respect to $\xi$, then take the limit $\xi \to 0$ of the primary antiderivative, thereby obtaining the UV-convergent primary antiderivative.\footnote{From the perspective of analytic continuation, the form of the primary antiderivative in the limit $\xi \to 0$ is fixed, allowing $\xi$ to be added to any denominator in a loop integral. This local outcome without UV divergences is the form we aim to obtain in loop calculations.} \\ (e) Set the boundary constant $C$ at specified reference energy points to derive the final result $\mathcal{T}_\mathrm{P}$.

\section{Applications}

Firstly, the above method is applied to specific processes as examples (see the Appendix \ref{add-ex} for additional examples). Then, we focus on the hierarchy problem of the Higgs mass.

\subsection{Some examples}

\subsubsection{The $\phi^4$ theory}

\begin{figure}[htbp!]
\includegraphics[width=0.44\textwidth]{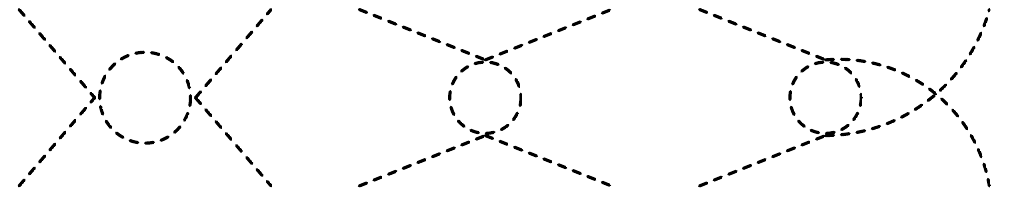} \vspace*{-1ex}
\caption{The one-loop diagrams of two-particle scatterings.}
\label{phy4}
\end{figure}

Let's apply the method to the $\phi^4$ theory. The Lagrangian of $\phi^4$ theory is
\begin{eqnarray}
\mathcal{L} = \frac{1}{2}(\partial_\mu \phi)^2  - \frac{1}{2} m^2 \phi^2 -  \frac{\lambda}{4!} \phi^4 \, .
\end{eqnarray}
The one-loop diagrams of two-particle scatterings in s, t and u channels are shown in Fig. \ref{phy4}, and the scattering amplitude has logarithmic divergences when evaluating loop integrations. Taking the approach described in Eq. (\ref{new-pro}), the Feynman-like scattering amplitude $\mathcal{T}_\mathrm{F} (\xi)$ in s channel can be written as
\begin{eqnarray}
\mathcal{T}_\mathrm{F} (\xi) \!=\! \frac{(-\mathrm{i} \lambda)^2}{2} \! \mathrm{\int} \! \frac{\mathrm{d}^Dk}{(2\pi)^D} \frac{\mathrm{i}}{k^2 \!-\!m^2 \!+ \!\xi} \frac{\mathrm{i}}{(k+q)^2 \!- \!m^2} \, ,  \label{xi-3}
\end{eqnarray}
where $q$ is the momentum transfer in the scattering process, with $q^2$ being equal to the Mandelstam $s$. The physical scattering amplitude $\mathcal{T}_\mathrm{P} (s)$ in this channel is
\begin{eqnarray} \label{s-channel}
\mathcal{T}_\mathrm{P} (s)\! & = & \! \Big [ \mathrm{\int} \mathrm{d}\xi \frac{\partial  \mathcal{T}_\mathrm{F} (\xi)}{\partial \xi} \Big]_{\xi \to 0} \! + \! C_1      \\
&=& \! \Big [ \! \frac{- \lambda^2}{2} \! \! \mathrm{\int} \mathrm{d}\xi  \! \mathrm{\int} \! \frac{\mathrm{d}^4k}{(2\pi)^4} \frac{- \mathrm{i}}{(k^2  \!- \!m^2  \!+ \!\xi)^2} \frac{\mathrm{i}}{(k \!+ \!q)^2  \!- \!m^2} \Big]_{\xi \to 0} \!   \nonumber \\
&& +   C_1 \, .         \nonumber
\end{eqnarray}
It is UV-converged when we evaluate the integration over the loop momentum $k$, and thus $D=4$ is adopted directly in the calculations of Eq. (\ref{s-channel}) (see the Appendix \ref{int-ex} for the transition from the traditional method to the UV-free scheme). After integral, one has
\begin{eqnarray}
\mathcal{T}_\mathrm{P} (s)\!
=  \frac{- \mathrm{i} \lambda^2}{32 \pi^2} \! \! \mathrm{\int}_{\! \! \! 0}^1 \mathrm{d} x  \log [m^2 - x (1-x)s ]+ \! C_1 \, . \label{phy-s}
\end{eqnarray}
Considering the renormalization conditions, the amplitudes are taken to be zero at $s = 4 m^2$, $t = u =0$. Thus, the constant $C_1$ here is
\begin{eqnarray}
C_1 =  \frac{ \mathrm{i} \lambda^2}{32 \pi^2} \! \! \mathrm{\int}_{\! \! \! 0}^1 \mathrm{d} x  \log [m^2 - 4 m^2 x (1-x) ] \, .
\end{eqnarray}

For t and u channels, similar results are obtained for $\mathcal{T}_\mathrm{P} (t)$ and $\mathcal{T}_\mathrm{P} (u)$, with $s$ in Eq. (\ref{phy-s}) replaced by $t$ and $u$ respectively. The total one-loop physical amplitude $\mathcal{T}_\mathrm{P}$ is
\begin{eqnarray}
\mathcal{T}_\mathrm{P} \! & = & \! \mathcal{T}_\mathrm{P} (s) +\mathcal{T}_\mathrm{P} (t) + \mathcal{T}_\mathrm{P} (u)     \\
&=& \! \frac{- \mathrm{i} \lambda^2}{32 \pi^2} \! \! \mathrm{\int}_{\! \! \! 0}^1 \mathrm{d} x  \Big [ \log   \frac{m^2 - x (1-x)s}{m^2 - 4 m^2 x (1-x)}    \nonumber \\
&& +  \log   \frac{m^2 - x (1-x)t}{m^2}     + \log   \frac{m^2 - x (1-x)u}{m^2}   \Big ]       \, .                                     \nonumber
\end{eqnarray}
We can see that the same finite result is obtained with the method here as dimensional regularization and renormalization, and there is no troublesome UV divergence in calculations. From another point of view, it gives an explanation why universal constant parts ($\gamma_E$, $\log(4 \pi)$) should be subtracted along with infinity in $\overline{\mathrm{MS}}$. In addition, one can check that the same result can be obtained if $\xi$ is added into the denominator of the second propagator in Eq. (\ref{xi-3}), and this is a freedom.

Here we consider the massless limit $m \to 0$. In this case, the one-loop correction to the two-point function is
\begin{eqnarray}
\mathcal{T}_\mathrm{P}^{2p}\! & = & \! \Big [ \mathrm{\int} \! \mathrm{d}\xi_1 \mathrm{d}\xi_2 \frac{\partial  \mathcal{T}_\mathrm{F}^{2p} (\xi_1,\xi_2)}{\partial \xi_1 \partial \xi_2} \Big]_{\{\xi_1, \xi_2\} \to 0} \! + \! C      \\
&=& \! \Big [ \! \frac{- \mathrm{i}\lambda}{2} \! \! \mathrm{\int} \! \mathrm{d}\xi_1 \mathrm{d}\xi_2  \!  \! \mathrm{\int} \! \! \frac{\mathrm{d}^4k}{(2\pi)^4} \frac{2\mathrm{i}}{(k^2 \!+ \!\xi_1 \!+ \!\xi_2)^3} \Big]_{\{\xi_1, \xi_2\} \to 0} \!  \!+   \! C   \, .                                               \nonumber
\end{eqnarray}
After integral, one has
\begin{eqnarray}
\mathcal{T}_\mathrm{P}^{2p}\!=\! \Big [ \! \frac{- \mathrm{i}\lambda}{2}  \frac{-1}{16\pi^2} (\log \xi_1^{\xi_1} - \xi_1) \Big]_{\xi_1 \to 0} \!+   \! C   \, .
\end{eqnarray}
The primary antiderivative is zero in the limit $\xi_1 \to 0$, and $C =$ 0 is adopted for this case. Thus, the one-loop correction to the two-point function is $\mathcal{T}_\mathrm{P}^{2p} =$ 0. In the massless limit, if the renormalization condition $s=-t=-u=\mu^2$ is adopted ($\mu$ is an energy scale), the total one-loop physical amplitude $\mathcal{T}_\mathrm{P}$ of the two-particle scatterings can be written as
\begin{eqnarray} \label{mu-phi}
\mathcal{T}_\mathrm{P} \! & = & \! \mathcal{T}_\mathrm{P} (s) +\mathcal{T}_\mathrm{P} (t) + \mathcal{T}_\mathrm{P} (u)     \\
&=& \! \frac{\mathrm{i} \lambda^2}{32 \pi^2}  \Big ( \log   \frac{\mu^2}{s} +\log   \frac{\mu^2}{-t} + \log   \frac{\mu^2}{-u}  \Big )       \, .    \nonumber
\end{eqnarray}

Now we turn to the behavior of the correlation function with the variation of the reference energy scale $\mu$. A physical quantity is considered to be independent of $\mu$ in the perturbation expansion. Let us take the electron physical charge $e = e_0 + \Delta e = e_\mu + \Delta e_\mu$ as an illustration, with $e_0$ being a bare parameter and $\Delta e$ being loop corrections, $e_\mu$ being the tree-level value at $\mu$ and $\Delta e_\mu$ being the corresponding loop corrections. In traditional renormalization, the finite $e$ is obtained via $\infty - \infty$, with both $e_0$ and $\Delta e$ being infinite (the top region). The physical structure of the traditional method is a top-down type, and the UV divergence $\infty$ in the top region is reduced to the low-energy region via $\infty - \infty$ ($e = e_0 + \Delta e$). In the view of UV-free scheme, the finite $e$ is obtained via two finite parameters $e_\mu$ and $\Delta e_\mu$ (the bottom region), i.e. a picture of locally finite quantities which are insensitive to the UV region. The physical structure of the UV-free scheme is a bottom-up type, with the UV region physics being free. In the UV-free scheme, the $n$-point physical correlation function $G^{(n)}_\mathrm{P}$ can be set by the physical field $\phi_\mathrm{P} (x)$ with $\phi_\mathrm{P} (x) = Z^{1/2} \phi (x, \mu)$, and the rescaling factor $Z$ is finite here. The local correlation function $G^{(n)}$ (shorthand for a full expression $G^{(n)} (\phi,\lambda,m,\cdots,\mu)$) at the scale $\mu$ can be written as $G^{(n)} = Z^{-n/2} G^{(n)}_\mathrm{P}$. Considering $\frac{\mathrm{d} G^{(n)}_\mathrm{P}}{\mathrm{d} \mu}=0$, the variation of $\mu$ in the massless limit can be described by a relation \begin{eqnarray}
(\mu \frac{\partial}{\partial \mu} + \beta \frac{\partial}{\partial \lambda} + n\gamma) G^{(n)} =0 \, .
\end{eqnarray}
This is the form of the Callan-Symanzik equation \cite{Callan:1970yg,Symanzik:1970rt}, and we have another picture about it in the UV-free scheme. In a loop correction, the explicit $\mu$-term is from the boundary constant $C$ at a reference energy scale $\mu$. For the $\phi^4$ theory in the massless limit, the one-loop result of the parameter $\gamma$ is zero ($\mathcal{T}_\mathrm{P}^{2p} =$ 0). The beta function can be derived by Eq. (\ref{mu-phi}), with
\begin{eqnarray}
\beta & = & -\mathrm{i} \mu \frac{\partial}{\partial \mu}  \mathcal{T}_\mathrm{P} \\
      & = &  \frac{3 \lambda^2}{16 \pi^2}   + \mathcal{O}  (\lambda^3)                     \, .     \nonumber
\end{eqnarray}
Being free of UV divergences, the running behavior can be reproduced in a locally finite picture that works well with the Callan-Symanzik equation. This is understandable, as a physical law can be expressed through several descriptive forms. Now let's see which are physical. The normalization conditions (renormalization conditions) and the running behavior are physical, while UV divergences $\infty$ arising in traditional technical methods appear to be intermediate variables.

\subsubsection{The axial anomaly}

\begin{figure}[htbp!]
\includegraphics[width=0.18\textwidth]{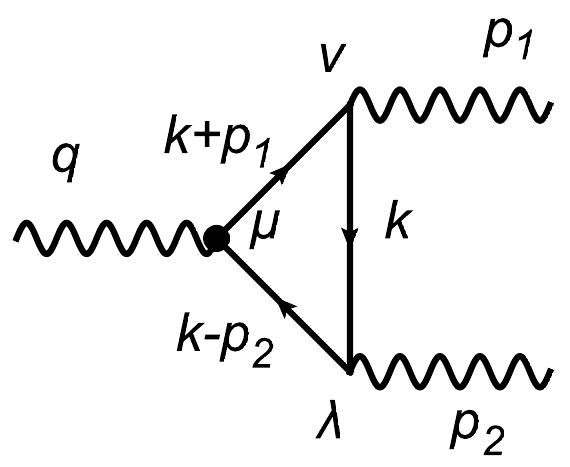} \vspace*{-1ex} \quad
\includegraphics[width=0.186\textwidth]{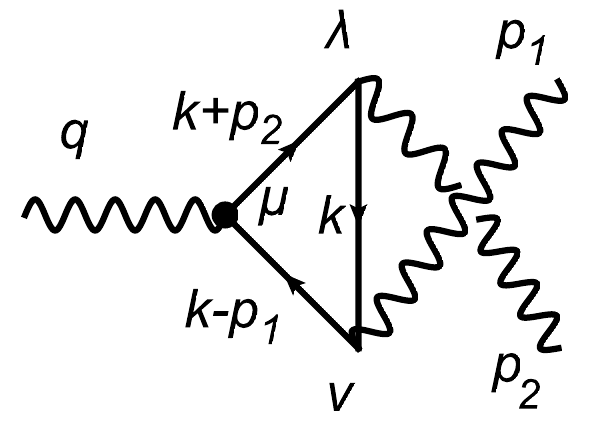} \vspace*{-1ex}
\caption{The one-loop diagrams of the axial vector current.}
\label{axial}
\end{figure}

The axial vector current $j^{\mu 5}$ is not conserved for massless fermions, with
\begin{eqnarray}
\partial_\mu j^{\mu 5} = -\frac{e^2}{16 \pi^2} \varepsilon^{\alpha \beta \mu \nu} F_{\alpha \beta } F_{\mu \nu} \, .   \label{axial-a}
\end{eqnarray}
This equation is the Adler-Bell-Jackiw anomaly \cite{Adler:1969gk,Bell:1969ts,Adler:1969er}. The axial anomaly can be checked by the transition of axial vector current $\to$ two photons being nonzero. The one-loop diagrams contributing to the two-photon matrix element of the axial vector current are shown in Fig. \ref{axial}. The physical transition amplitude $\mathcal{T}_\mathrm{P}^{\mu \nu \lambda}$ to the axial current can be written as
\begin{eqnarray}
\mathrm{i} q_\mu \mathcal{T}_\mathrm{P}^{\mu \nu \lambda} \!&=&\! \mathrm{i} q_\mu \! \Big( \Big[ \mathrm{\int} \mathrm{d}\xi_1 \frac{\partial  \mathcal{T}_\mathrm{F}^{\mu \nu \lambda} (\xi_1)}{\partial \xi_1} \Big]_{\xi_1 \to 0} \! + \! C_1^{\mu \nu \lambda}   \nonumber \\
&&    + [ \nu \leftrightarrow \lambda , p_1 \leftrightarrow p_2] \Big)     \, \\
&=&  \mathrm{i} q_\mu (-\mathrm{i} e)^2 (-\mathrm{i}) \Big( \Big[ \mathrm{\int} \mathrm{d}\xi_1 \! \mathrm{\int} \! \frac{\mathrm{d}^4k}{(2\pi)^4} \!   \nonumber \\
&& \!\!\!\!\! \times \mathrm{tr} \big( \gamma^\mu \gamma^5  \! \frac{\slashed{k}-\slashed{p}_2}{((k-p_2)^2  \!+ \!\xi_1)^2} \gamma^\lambda \! \frac{\slashed{k}}{k^2}  \gamma^\nu \!\frac{\slashed{k}+\slashed{p}_1}{(k \!+ \!p_1)^2 } \big) \Big]_{\xi_1 \to 0}                                 \nonumber \\
&& + C_1^{\mu \nu \lambda} + [ \nu \leftrightarrow \lambda , p_1 \leftrightarrow p_2] \Big)  \, . \nonumber
\end{eqnarray}
Taking the trace of $\gamma-$matrices and evaluating the integrations, one has
\begin{eqnarray}
\mathrm{i} q_\mu \mathcal{T}_\mathrm{P}^{\mu \nu \lambda} \!&=&\! \frac{(-\mathrm{i} e)^2}{4 \pi^2}\Big( \! \mathrm{\int}_{\! \! \! 0}^1 \!  \mathrm{d}x_1 \mathrm{d}x_2 \mathrm{d}x_3  \delta(1\!-\!x_1\!-\!x_2\!-\!x_3)  \nonumber  \\
&& \times  \big[ 6 (1-\frac{x_1 + x_3}{2})\log\frac{1}{2 x_1 x_3 p_1\cdot p_2}  \nonumber \\
&& + (x_1 + x_3 -2)  + C_1 \big] \varepsilon^{\alpha \lambda \beta \nu}{p_1}_\alpha {p_2}_\beta \nonumber  \\
&& + [ \nu \leftrightarrow \lambda ,  p_1 \leftrightarrow p_2] \Big)  \,   \\
&=&\frac{(-\mathrm{i} e)^2}{4 \pi^2} \! \mathrm{\int}_{\! \! \! 0}^1 \!  \mathrm{d}x_1 \mathrm{d}x_2 \mathrm{d}x_3  \delta(1\!-\!x_1\!-\!x_2\!-\!x_3) \nonumber  \\
&& \times \big[ 6 (1-\frac{x_1 + x_3}{2})\log\frac{1}{2 x_1 x_3 p_1\cdot p_2}  \nonumber \\
&& +(x_1 + x_3 -2) + C_1 \big] 2 \varepsilon^{\alpha \lambda \beta \nu}{p_1}_\alpha {p_2}_\beta \nonumber \, .
\end{eqnarray}
The term $(x_1 + x_3 -2)$ is originally finite, and the Ward identity is automatically preserved. Suppose $C_1$ is related to an energy scale $\mu^2 = r \times 2 x_1 x_3 p_1 \cdot p_2 $ with $r$ being a relative coefficient, and the result can be written as
\begin{eqnarray}
\mathrm{i} q_\mu \mathcal{T}_\mathrm{P}^{\mu \nu \lambda} \!&=&\! \frac{(-\mathrm{i} e)^2}{4 \pi^2} \! \mathrm{\int}_{\! \! \! 0}^1 \!  \mathrm{d}x_1 \mathrm{d}x_2 \mathrm{d}x_3  \delta(1\!-\!x_1\!-\!x_2\!-\!x_3)  \\
&& \times \big[(x_1 + x_3 -2) (1-3\log r)\big] 2 \varepsilon^{\alpha \lambda \beta \nu}{p_1}_\alpha {p_2}_\beta  \nonumber \\
&=& - \frac{(-\mathrm{i} e)^2}{2 \pi^2}  (\frac{2}{3}-2\log r)   \varepsilon^{\alpha \lambda \beta \nu}{p_1}_\alpha {p_2}_\beta \nonumber         \, .
\end{eqnarray}
Now, the result is
\begin{eqnarray}
\partial_\mu j^{\mu 5} &=& \mathrm{i} q_\mu \mathcal{T}_\mathrm{P}^{\mu \nu \lambda} \epsilon_\nu^\ast (p_1)  \epsilon_\lambda^\ast (p_2)
 \\
&=&  -\frac{e^2}{16 \pi^2} (\frac{2}{3}-2\log r) \varepsilon^{\alpha \nu \beta \lambda} F_{\alpha \nu} F_{\beta \lambda}     \nonumber \, .
\end{eqnarray}
This result is consistent with the structure of Eq. (\ref{axial-a}). The leading order here arises from loop diagrams, in contrast to cases dominated by tree-level processes. Thus, apart from a reference energy scale $\mu$, there is currently a lack of more appropriate conditions to further set the value of it. Taking $C_0 = 2\log r$, the value of $C_0$ is of order one estimated by naturalness. If Eq. (\ref{axial-a}) is considered as a relation that the axial vector current should follow, the value $C_0 = -\frac{1}{3}$ is obtained with SM being a self-consistent theory. Moreover, the values $\frac{2}{3}$, $-\frac{1}{3}$ are equal to the charge values of quarks, and it is not known whether it is a coincidence or there may be some correlation between them. From this perspective, the appropriate conditions for the axial anomaly will hold profound physical significance.

\subsection{The hierarchy problem}

\begin{figure}[htbp!]
\includegraphics[width=0.465\textwidth]{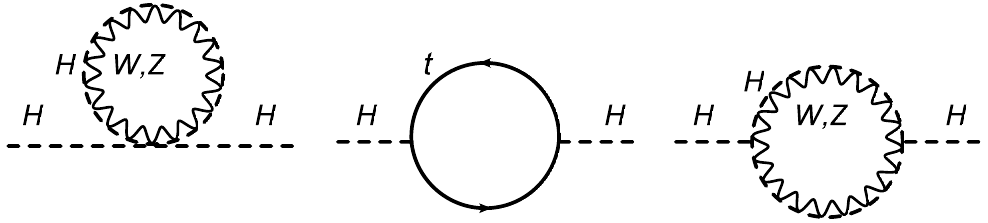} \vspace*{-1ex}
\caption{The one-loop corrections to the Higgs boson mass.}
\label{higgs1}
\end{figure}

With the discovery of the Higgs boson at LHC \cite{ATLAS:2012yve,CMS:2012qbp}, the Higgs boson that is not too heavy (125 GeV) accentuates the hierarchy problem, i.e. the naturalness of the fine-tuning originating from the radiative corrections to the Higgs mass. The one-loop radiative corrections to the Higgs mass are power-law divergences, as depicted in Fig. \ref{higgs1}. What prevents the Higgs mass getting quantum corrections from very high energy scale (the Planck scale)? At present, it is still an open question. Here we try to give it an answer in the UV-free scheme.

The radiative corrections from the Higgs boson in the first diagram of Fig. \ref{higgs1} is
\begin{eqnarray}
\mathcal{T}_\mathrm{P}^{H1} \! & = & \! \Big [  \mathrm{\int} \mathrm{d}\xi_1 \mathrm{d}\xi_2 \frac{\partial  \mathcal{T}_\mathrm{F}^{H1} (\xi_1, \xi_2 )}{\partial \xi_1 \partial \xi_2} \Big]_{\{\xi_1, \xi_2\} \to 0} \! + \! C     \nonumber  \\
&=& \! \Big [  (- 3\mathrm{i})\! \frac{m_H^2}{2 v^2} \! \! \mathrm{\int}  \mathrm{d}\xi_1 \mathrm{d}\xi_2   \! \mathrm{\int} \! \frac{\mathrm{d}^4k}{(2\pi)^4} \!   \\
&& \times \frac{2 \mathrm{i} }{(k^2 -m_H^2 +\xi_1 +\xi_2)^3} \Big]_{\{\xi_1, \xi_2\} \to 0} \! +  C\,    .              \nonumber
\end{eqnarray}
The primary antiderivative is composed solely of the $\xi$ terms. After integral, one has
\begin{eqnarray}
\mathcal{T}_\mathrm{P}^{H1} \!  &=& \mathrm{i} \frac{3 m_H^4}{32 \pi^2 v^2} \! \log \frac{1}{m_H^2}   +  C      \\
&=& \mathrm{i} \frac{3 m_H^4}{32 \pi^2 v^2} \! \log \frac{\mu^2}{m_H^2}        \,    .              \nonumber            \end{eqnarray}

Now we turn to the loop of vector boson V (V=W,Z) shown in the first diagram of Fig. \ref{higgs1}. In unitary gauge, the corresponding superficial degree of divergence is increased to 4. The radiative corrections with these quartic divergences can be calculated, with
\begin{eqnarray}
\mathcal{T}_\mathrm{P}^{V1} \! & = & \! \Big [  \mathrm{\int} \mathrm{d}\xi_1 \mathrm{d}\xi_2 \mathrm{d}\xi_3 \frac{\partial  \mathcal{T}_\mathrm{F}^{V1} (\xi_1, \xi_2, \xi_3)}{\partial \xi_1 \partial \xi_2  \partial \xi_3} \Big]_{\{\xi_1, \xi_2, \xi_3\} \to 0} \! + \! C     \nonumber  \\
&=& \! \Big [  \mathrm{i}  \frac{2 m_V^2}{v^2 s_V} \! \! \mathrm{\int}  \mathrm{d}\xi_1 \mathrm{d}\xi_2 \mathrm{d}\xi_3  \! \mathrm{\int} \! \frac{\mathrm{d}^4k}{(2\pi)^4}    g_{\mu \nu} \\
&& \times \frac{6 \mathrm{i} (g^{\mu \nu}- k^\mu k^\nu/m_V^2) }{(k^2 -m_V^2 +\xi_1 +\xi_2 +\xi_3)^4} \Big]_{\{\xi_1, \xi_2, \xi_3 \} \to 0} \! +  C\,  \,    ,             \nonumber
\end{eqnarray}
where the symmetry factor $s_V$ is $s_V=$ 1, 2 for W, Z respectively. After integral, one has
\begin{eqnarray}
\mathcal{T}_\mathrm{P}^{V1} \! &=& \!  \mathrm{i}  \frac{2 m_V^2}{v^2 s_V}  \frac{m_V^2}{16 \pi^2} (3 \log \frac{1}{m_V^2})\! +  C\,  \,  \\
 &=& \!  \mathrm{i}  \frac{2 m_V^2}{v^2 s_V}  \frac{3 m_V^2}{16 \pi^2}  \log \frac{\mu^2}{m_V^2} \,  \, .  \nonumber
\end{eqnarray}

The top quark loop is shown in the second diagram of Fig. \ref{higgs1}, and the corresponding radiative correction is
\begin{eqnarray}
\mathcal{T}_\mathrm{P}^{\, t} \! & = & \! \Big [  \mathrm{\int} \mathrm{d}\xi_1 \mathrm{d}\xi_2 \frac{\partial  \mathcal{T}_\mathrm{F}^{\, t} (\xi_1, \xi_2 )}{\partial \xi_1 \partial \xi_2} \Big]_{\{\xi_1, \xi_2\} \to 0} \! + \! C     \nonumber  \\
&=& \! \Big [ \frac{3 m_t^2}{ v^2} \! \! \mathrm{\int}  \mathrm{d}\xi_1 \mathrm{d}\xi_2   \! \mathrm{\int} \! \frac{\mathrm{d}^4k}{(2\pi)^4} \!   \\
&& \!\!\! \times \mathrm{tr} \big(  \frac{2 \mathrm{i} (\slashed{k}+m_t)}{(k^2 \! - \!m_t^2 \!+\! \xi_1 \!+\! \xi_2)^3}  \frac{\mathrm{i}(\slashed{p}+\slashed{k}+m_t)}{(p+k)^2 \!-\! m_t^2} \big) \Big]_{\{\xi_1, \xi_2\} \to 0} \! +\!  C\,    ,\nonumber
\end{eqnarray}
with the external momentum $p$. After integral, one has
\begin{eqnarray}
\mathcal{T}_\mathrm{P}^{\, t} \! &=& \!  -  \frac{3 m_t^2}{ v^2} \!\frac{\mathrm{i}}{4 \pi^2}  \!\mathrm{\int}_{\! \! \! 0}^1 \mathrm{d} x  [m_t^2 - p^2 x(1-x)] \\
&& \times (3\log\frac{1}{m_t^2 - p^2 x(1-x)}) +  C   \nonumber \\
&=& \!  -  \frac{3 m_t^4}{ v^2} \!\frac{3\mathrm{i}}{4 \pi^2}  \!\mathrm{\int}_{\! \! \! 0}^1 \mathrm{d} x  [1 - \frac{p^2}{m_t^2} x(1-x)]     \nonumber  \\
&&  \times \log\frac{\mu^2}{m_t^2 - p^2 x(1-x)}  \,  .              \nonumber
\end{eqnarray}

The radiative correction of Higgs loop shown in the third diagram of Fig. \ref{higgs1} is
\begin{eqnarray}
\mathcal{T}_\mathrm{P}^{H3} \! & = & \! \Big [  \mathrm{\int} \mathrm{d}\xi_1 \frac{\partial  \mathcal{T}_\mathrm{F}^{H3} (\xi_1  )}{\partial \xi_1} \Big]_{\xi_1 \to 0} \! + \! C     \\
&=& \! \Big [  (- 3\mathrm{i})^2  \frac{m_H^4}{2 v^2} \! \! \mathrm{\int}  \mathrm{d}\xi_1    \! \mathrm{\int} \! \frac{\mathrm{d}^4k}{(2\pi)^4} \!     \nonumber   \\
&& \times \frac{- \mathrm{i}}{(k^2 -m_H^2 +\xi_1)^2} \frac{\mathrm{i}}{(k+p)^2 -m^2} \Big]_{\xi_1 \to 0} \! +  C\,    .              \nonumber
\end{eqnarray}
After integral, one has
\begin{eqnarray}
\mathcal{T}_\mathrm{P}^{H3} &=& \frac{9 m_H^4}{2 v^2} \frac{\mathrm{i}}{16 \pi^2} \mathrm{\int}^1_{\!\!\! 0} \mathrm{d} x \, \log\frac{1}{m_H^2 - x (1-x)p^2} +C   \nonumber \\
&=& \mathrm{i} \frac{9 m_H^4}{32 \pi^2 v^2} \mathrm{\int}^1_{\!\!\! 0} \mathrm{d} x \, \log\frac{\mu^2}{m_H^2 - x (1-x)p^2} \, .
\end{eqnarray}
The radiative correction of vector boson V loop shown in the third diagram of Fig. \ref{higgs1} is
\begin{eqnarray}
\mathcal{T}_\mathrm{P}^{V3} \! & = & \! \Big [  \mathrm{\int} \mathrm{d}\xi_1 \mathrm{d}\xi_2 \mathrm{d}\xi_3 \frac{\partial  \mathcal{T}_\mathrm{F}^{V3} (\xi_1, \xi_2, \xi_3)}{\partial \xi_1 \partial \xi_2  \partial \xi_3} \Big]_{\{\xi_1, \xi_2, \xi_3\} \to 0} \! + \! C      \\
&=& \! \Big [ \! - \! \frac{4 m_V^4}{v^2 s_V} \! \! \mathrm{\int} \! \mathrm{d}\xi_1 \mathrm{d}\xi_2 \mathrm{d}\xi_3  \! \mathrm{\int} \! \frac{\mathrm{d}^4k}{(2\pi)^4}   \frac{6 \mathrm{i} (g^{\mu \nu}- k^\mu k^\nu/m_V^2) }{(k^2 \!-\! m_V^2 \!+\! \xi_1 \!+\! \xi_2 \!+\! \xi_3)^4}  \nonumber \\
&& \times \frac{-\mathrm{i}(g_{\mu \nu} \!-\! (k\!+\!p)_\mu (k\!+\!p)_\nu/m_V^2)}{(k+p)^2 -m_V^2} \Big]_{\{\xi_1, \xi_2, \xi_3 \} \to 0} \!+\!  C   \,    .   \nonumber
\end{eqnarray}
After integral, one has
\begin{eqnarray}
\mathcal{T}_\mathrm{P}^{V3} \! &=& \! \frac{4 m_V^4}{v^2 s_V} \frac{6\mathrm{i}}{16 \pi^2} \mathrm{\int}^1_{\!\!\! 0} \mathrm{d} x  \big[ \frac{1}{2} - \frac{p^2}{m_V^2}(x-x^2+\frac{1}{12})              \\
&&  + \frac{p^4}{m_V^4}\frac{x (1\!-\!x)}{12} (20x\!-\!20x^2\!-\!1)\big]  \nonumber\\
&& \times \log\frac{1}{m_V^2 \!-\! x (1\!-\!x)p^2} \!+\!  C    \nonumber \\
&=& \!  \frac{ m_V^4}{v^2 s_V} \frac{3\mathrm{i}}{2 \pi^2} \mathrm{\int}^1_{\!\!\! 0} \mathrm{d} x  \big[ \frac{1}{2} \!-\! \frac{p^2}{m_V^2}(x-x^2+\frac{1}{12})        \nonumber      \\
&& \!\!\! + \frac{p^4}{m_V^4}\frac{x (1\!-\!x)(20x\!-\!20x^2\!\!-\!1)}{12} \big]\!\log\frac{\mu^2}{m_V^2 \!\!-\! x (1\!-\!x)p^2}     \,  .   \nonumber
\end{eqnarray}

Considering the reference energy scale $\mu$ in the electroweak scale, the above corrections (multiplied by $\mathrm{i}$) to the Higgs mass without fine-tuning are not very large. Moreover, if on-shell renormalization conditions are adopted, the results can be written as
\begin{eqnarray}
\mathcal{T}_\mathrm{P}^{H1} \!  = \mathcal{T}_\mathrm{P}^{V1} = 0      \,    ,
\end{eqnarray}
\begin{eqnarray}
\mathcal{T}_\mathrm{P}^{\, t} \! &=& -  \frac{3 m_t^4}{ v^2} \!\frac{3\mathrm{i}}{4 \pi^2}  \!\mathrm{\int}_{\! \! \! 0}^1 \mathrm{d} x  [1 - \frac{p^2}{m_t^2} x(1-x)]       \\
&& \times \log\frac{m_t^2 - m_H^2 x(1-x)}{m_t^2 - p^2 x(1-x)}  \,  ,             \nonumber
\end{eqnarray}
\begin{eqnarray}
\mathcal{T}_\mathrm{P}^{H3} = \mathrm{i} \frac{9 m_H^4}{32 \pi^2 v^2} \mathrm{\int}^1_{\!\!\! 0} \mathrm{d} x \, \log\frac{m_H^2 - m_H^2 x (1-x)}{m_H^2 - x (1-x)p^2} \, ,
\end{eqnarray}
\begin{eqnarray}
\mathcal{T}_\mathrm{P}^{V3} \! &=& \! \frac{ m_V^4}{v^2 s_V} \frac{3\mathrm{i}}{2 \pi^2} \mathrm{\int}^1_{\!\!\! 0} \mathrm{d} x  \big[ \frac{1}{2} - \frac{p^2}{m_V^2}(x-x^2+\frac{1}{12})              \\
&& \!\!\!\! + \frac{p^4}{m_V^4}\frac{x (1\!-\!x)(20x\!-\!20x^2\!\!-\!1)}{12} \big]\!\log \!\frac{m_V^2 \!\!-\! x (1\!-\!x)m_H^2}{m_V^2 \!\!-\! x (1\!-\!x)p^2}    \,  .   \nonumber
\end{eqnarray}
Loop corrections to the Higgs mass are tangible in UV-free scheme, and the Higgs boson with 125 GeV mass can be obtained without fine-tuning. Thus, it gives an interpretation to the hierarchy problem within SM.

Here we give a brief discussion about loops with different methods. The usual procedure for UV divergences of loops is regularization and renormalization, and this paradigm is based on the Bogoliubov-Parasiuk-Hepp-Zimmermann (BPHZ) renormalization scheme \cite{Bogoliubov:1957gp}, i.e., all UV divergences can be removed by a finite number of counterterms in a renormalizable theory. In this paper, a method of the UV-free scheme described by Eq. (\ref{new-pro}) is introduced to obtain loop results. Let's look at these two ways from the perspective of divergences. For logarithmic divergences, both a suitable regulator with the BPHZ scheme and Eq. (\ref{new-pro}) can cure UV divergences and derive the finite loop results. For power-law divergences such as loop corrections of the Higgs mass, the loop results obtained via renormalization are fine-tuned \cite{Branchina:2022jqc,Branchina:2022gll}, while in the UV-free scheme, finite loop results can be obtained without fine-tuning. Why does the UV-free scheme seem effective and still hold for power-law divergences? This is due to two different routes adopted in loop calculations:\\ (a) \textit{Equivalent transformation} of the loop integral from UV divergence to UV divergence mathematically expressed form (regularization), with renormalization required to remove the UV divergence ($\infty - \infty$). \\
(b) \textit{Analytic continuation} of the transition amplitude from UV-divergent $\mathcal{T}_\mathrm{F}$ written by Feynman rules to UV-converged $\mathcal{T}_\mathrm{P}$ ($\mathcal{T}_\mathrm{F} \to \mathcal{T}_\mathrm{P}$, the UV-free scheme here), with UV divergences systematically eliminated. \\
Another interpretation may be necessary, that is, the traditional methods of the Pauli-Villars regularization and the derivative method set out from the propagator, while these two methods go along the route of equivalence transformation and the final results are derived via $\infty - \infty$. Being free from possible troubles caused by $\infty - \infty$, the loop result obtained by the UV-free scheme via the analytic continuation $\mathcal{T}_\mathrm{F} \to \mathcal{T}_\mathrm{P}$ continues to work for power-law divergences.

Additionally, here we give an explanation for why the parameter $\xi$ needs to be introduced in the propagator rather than being directly applied to the external momentum and mass. For instance, in gluon seagull diagram and one-loop propagator in massless $\phi^4$ theory, the loop integral neither involves external momentum nor contains mass term, making the parameter $\xi$ particularly essential in such cases (for these two diagrams, the loop contributions in the UV-free scheme are zero). Therefore, the parameter $\xi$ can provide a unified description.

\section{Conclusion and Discussion}
\label{sec:Con}

The UV divergences of loops with finite physical results obtained via $\infty - \infty$ indicate that transition amplitudes directly obtained are not always well-defined, as pointed out by Dirac \cite{DiracTheEO} and Feynman \cite{Feynman:1986er}. If we go forward, the transition amplitude directly obtained by Feynman rules is taken as physical input, and the physical result is taken as physical output. Thus, the physical output depends on the physical input, but not directly equals to the physical input. In this paper, a presumption of the physical contributions of loops from UV regions being insignificant is proposed, and we find that the physical output can be described by Eq. (\ref{new-pro}) with a method of UV-free scheme, i.e. loop results can be obtained without UV divergences. When performing $\xi$ changes on a gauge field propagator, the gauge invariance can be considered as the physical input required, or being formally restored after taking the antiderivative with respect to $\xi$. The UV-free scheme takes a route of analytic continuation between the physical input and output:
\begin{eqnarray}
 \mathrm{Input} \bigg \{\! \begin{array}{c}
  (a) \textit{Equivalent transformation} \, , \infty - \infty\\
  (b) \textit{Analytic continuation} \, , ~ \mathcal{T}_\mathrm{F} \to \mathcal{T}_\mathrm{P}~~~
\end{array} \!\bigg \} \mathrm{Output} \,. \nonumber
\end{eqnarray}
This method is effective for both trivial cases (tree level and originally finite loops) and UV-divergent cases (loops with logarithmic and power-law divergences), which indicates that the new method has a logical consistency in describing the transition amplitude.

To the open question of the hierarchy problem of Higgs mass with power-law divergences, in the UV-free scheme, loop corrections to Higgs mass are not very large, and a 125 GeV Higgs can be obtained without fine-tuning, i.e. an interpretation within SM. Moreover, if SM is considered as an effective field theory at low energy scale, loop corrections in low-energy processes are insensitive to the contributions of loop momenta from very high energy scale (e.g. the Planck scale), with possible new physics in the UV regions being free. This is the locality of loop corrections in the low-energy region. Besides the applications in SM, the method may also be capable to describe loops of the non-renormalizable field of Einstein gravity (see the companion article \cite{Jia:2024lyy}). Now, a method of UV-free scheme has been preliminarily introduced on loops, and we hope it is the beginning of a new method and look forward to more explorations.

\acknowledgments

The author thanks Yan-Qing Ma for helpful discussions. This work was partly supported by the open project of the theoretical physics academic exchange platform of Chongqing University.

\appendix

\section{Additional Examples}  \label{add-ex}

\subsubsection{The gauge field propagator}

If the method is applied to a gauge field propagator without free loop momentum, e.g., the photon propagator $ \frac{-\mathrm{i} g_{\mu \nu}}{p^2 +\mathrm{i} \epsilon}$, the result can be written as $\mathcal{T}_{\mathrm{F}} (\xi)\!=\!\frac{-\mathrm{i} g_{\mu \nu}}{p^2 + \xi + \mathrm{i} \epsilon}$, \!$\frac{\partial  \mathcal{T}_{\mathrm{F}} (\xi)}{\partial \xi}\!=\!\frac{-\mathrm{i} g_{\mu \nu}(-1)}{(p^2 + \xi + \mathrm{i} \epsilon)^2}$, \!$\big[\int \! \mathrm{d} \xi \frac{\partial  \mathcal{T}_{\mathrm{F}} (\xi)}{\partial \xi} \big]\!=\!\frac{-\mathrm{i} g_{\mu \nu}}{p^2 + \xi + \mathrm{i} \epsilon}$, with the boundary constant $C =$ 0 adopted without free loop momentum. The final result is $\big [\! \int \! \mathrm{d} \xi \frac{\partial  \mathcal{T}_{\mathrm{F}} (\xi)}{\partial \xi} \! \big]_{\xi \to 0}$ = $\frac{-\mathrm{i} g_{\mu \nu}}{p^2 +\mathrm{i} \epsilon}$, with the gauge field propagator restored.

\subsubsection{The electron self-energy}

\begin{figure}[htbp!]
\includegraphics[width=0.3\textwidth]{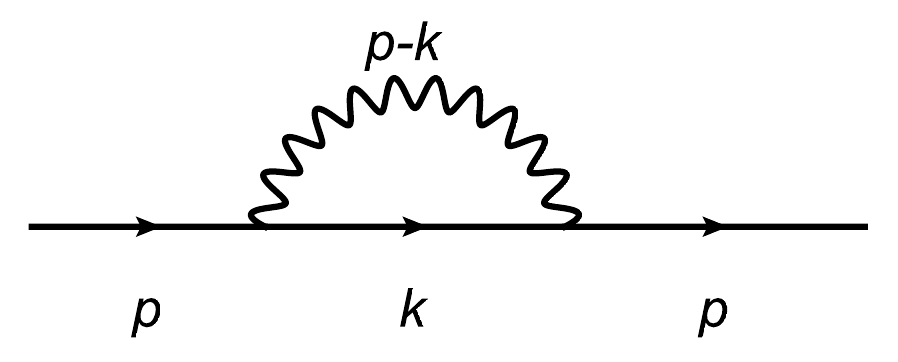} \vspace*{-1ex}
\caption{The one-loop diagram of electron self-energy.}
\label{elec-s}
\end{figure}

Now, we turn to the electron self-energy. The one-loop diagram is shown in Fig. \ref{elec-s}, and the transition amplitude has logarithmic divergence when evaluating the loop integration. The physical transition amplitude $\mathcal{T}_\mathrm{P}$  is
\begin{eqnarray}
\mathcal{T}_\mathrm{P} \! & = & \! \Big [ \mathrm{\int} \mathrm{d}\xi \frac{\partial  \mathcal{T}_\mathrm{F} (\xi)}{\partial \xi} \Big]_{\xi \to 0} \! + \! C      \\
&=& \! \Big [ \! (- \mathrm{i} e)^2 \! \! \mathrm{\int} \mathrm{d}\xi  \! \mathrm{\int} \! \frac{\mathrm{d}^4k}{(2\pi)^4}\gamma^\mu \frac{-\mathrm{i} (\slashed{k}+m)}{(k^2 \!-\!m^2 \!+\!\xi \!+\! \mathrm{i}\epsilon)^2} \gamma_\mu     \nonumber \\
&& \times \frac{-\mathrm{i}}{(p\!-\!k)^2\!+\! \mathrm{i}\epsilon} \Big]_{\xi \to 0} \! +   C \, .              \nonumber
\end{eqnarray}
After integral, one has
\begin{eqnarray}
\!\!\!\!\! \mathcal{T}_\mathrm{P} \! & =& \! - \mathrm{i}\frac{\alpha}{2 \pi} \!\! \mathrm{\int}_{\! \! \! 0}^1 \!\! \mathrm{d} x (2m\!-\!x\slashed{p})\! \log \! \frac{1}{(1\!-\!x)(m^2 \!\!-\! x p^2)} \!+\!C  .
\end{eqnarray}
If a reference energy scale $\mu$ is adopted with $C$ absorbed into the log term to make it dimensionless, the result is
\begin{eqnarray}
\!\! \mathcal{T}_\mathrm{P} \!  = \!  - \mathrm{i}\frac{\alpha}{2 \pi} \! \mathrm{\int}_{\! \! \! 0}^1 \!\! \mathrm{d} x (2m\!-\!x\slashed{p}) \! \log \! \frac{\mu^2}{(1\!-\!x)(m^2 \!-\! x p^2)}\, .
\end{eqnarray}
If the on-shell renormalization conditions are adopted for this process, the result is
\begin{eqnarray}
\mathcal{T}_\mathrm{P} \!  = \!  - \mathrm{i}\frac{\alpha}{2 \pi} \! \mathrm{\int}_{\! \! \! 0}^1 \mathrm{d} x (2m-x\slashed{p}) \! \log \! \frac{(1-x) m^2}{m^2 - x p^2}\, .
\end{eqnarray}

\subsubsection{A two-loop example}

\begin{figure}[htbp!]
\includegraphics[width=0.15\textwidth]{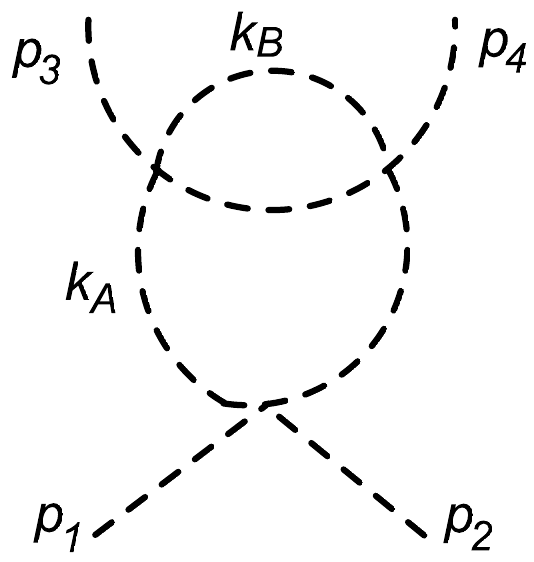} \vspace*{-1ex}
\caption{A two-loop transition.}
\label{tl}
\end{figure}

Here the method is applied to a two-loop transition with overlapping divergences in the $\phi^4$ theory, as shown in Fig. \ref{tl}. There are two-free loop momenta $k_A$ and $k_B$, and the physical transition amplitude $\mathcal{T}_\mathrm{P}$  is
\begin{eqnarray}
\mathcal{T}_\mathrm{P} \! & = & \! \Big [ \mathrm{\int} \mathrm{d}\xi \frac{\partial  \mathcal{T}_\mathrm{F} (\xi)}{\partial \xi} \Big]_{\xi \to 0} \! + \! C      \\
&=& \! \Big [ \! \frac{(- \mathrm{i}\lambda)^3}{2} \! \! \mathrm{\int} \!   \mathrm{d}\xi \! \mathrm{\int} \! \! \frac{\mathrm{d}^4k_A}{(2\pi)^4} \! \frac{\mathrm{d}^4k_B}{(2\pi)^4} \frac{\mathrm{ i}}{k_A^2  \!- \!m^2 } \frac{\mathrm{i}}{(k_A \!\!+ \!q)^2 \! \!- \!m^2} \nonumber \\
&& \times  \! \frac{-\mathrm{ i}}{(k_B^2  \!- \!m^2  \!+ \!\xi)^2} \frac{\mathrm{i}}{(k_B +k_A \!+ \!p_3)^2  \!- \!m^2} \Big]_{\xi \to 0} \! + \! C \nonumber   \, ,
\end{eqnarray}
with $q\!=\!p_1\!+\!p_2$. After evaluating the integration over $k_B$, one has
\begin{eqnarray}
\mathcal{T}_\mathrm{P} \! &=& \! \Big [ \! \frac{(- \mathrm{i}\lambda)^3}{2} \! \!\mathrm{\int}^1_{\!\!\! 0} \! \mathrm{d} x  \! \mathrm{\int} \!  \mathrm{d}\xi     \! \! \mathrm{\int} \! \! \frac{\mathrm{d}^4k_A}{(2\pi)^4}  \! \frac{\mathrm{i}}{k_A^2  \!\!- \!m^2} \frac{\mathrm{i}}{(k_A \!\!+ \!q)^2 \! \!- \!m^2}   \nonumber  \\
&& \!\!\!\!\!\! \times  \!  \frac{x }{16 \pi^2} \! \frac{\mathrm{i}}{(k_A \!+ \!p_3)^2 x(1\!-\!x)\!-\!m^2\!+\!x\xi}\!  \Big]_{\xi \to 0} \! \!+\! C  \,  .
\end{eqnarray}
The expression can be rewritten as
\begin{eqnarray}
\!\! \mathcal{T}_\mathrm{P} \! &=& \! \Big [ \! \frac{(- \mathrm{i}\lambda)^3}{2} \! \!\mathrm{\int}^1_{\!\!\! 0} \! \mathrm{d} x  \!\mathrm{\int}^1_{\!\!\! 0} \! \mathrm{d} y  \! \mathrm{\int} \!  \mathrm{d}\xi   \! \! \mathrm{\int} \! \! \frac{\mathrm{d}^4k_A}{(2\pi)^4}  \! \frac{-1}{(k_A^2 \!+\!2y k_A\!\cdot\!q \!+\!y q^2\!-\!m^2)^2}   \nonumber  \\
&&  \times  \!  \frac{x }{16 \pi^2} \! \frac{\mathrm{i}}{(k_A \!+ \!p_3)^2 x(1\!-\!x)\!-\!m^2\!+\!x\xi}\!  \Big]_{\xi \to 0} \! + \! C       \\
&=& \! \Big [ \! \frac{(- \mathrm{i}\lambda)^3}{2} \! \!\mathrm{\int}^1_{\!\!\! 0} \! \mathrm{d} x  \!\mathrm{\int}^1_{\!\!\! 0} \! \mathrm{d} y  \!\mathrm{\int}^1_{\!\!\! 0} \! \mathrm{d} z  \!  \mathrm{\int} \!  \mathrm{d}\xi   \! \! \mathrm{\int} \! \! \frac{\mathrm{d}^4k_A}{(2\pi)^4}  \!\frac{- \mathrm{i} }{16 \pi^2 (1\!-\!x)}    \nonumber  \\
&&  \times  \! \frac{2(1-z)}{[z D_B^{}\!+\!(1\!-\!z)D_A^{}]^3}\!  \Big]_{\xi \to 0} \! + \! C    \, ,   \nonumber
\end{eqnarray}
with $D_A^{}\!=\!k_A^2 \!+\!2y k_A\!\cdot\!q \!+\!y q^2\!-\!m^2$, $D_B^{}\!=\!(k_A \!+\!p_3)^2 \!-\!m^2/x(1\!-\!x)\!+\!\xi/(1\!-\!x)$. After the $k_A$ integral, one has
\begin{eqnarray}
\!\!\! \mathcal{T}_\mathrm{P} \! &=& \! \Big [ \! \frac{(- \mathrm{i}\lambda)^3}{2} \! \!\mathrm{\int}^1_{\!\!\! 0} \! \mathrm{d} x  \!\mathrm{\int}^1_{\!\!\! 0} \! \mathrm{d} y  \!\mathrm{\int}^1_{\!\!\! 0} \! \mathrm{d} z  \!  \mathrm{\int} \!  \mathrm{d}\xi     \!\frac{- \mathrm{i}x }{16 \pi^2}    \! \frac{-\mathrm{i}(1-z)}{16 \pi^2 (\Delta\!-\!xz\xi)}\!  \Big]_{\xi \to 0} \! + \! C  \nonumber \\
&=& \! \frac{(- \mathrm{i}\lambda)^3}{2 (4\pi)^4} \! \!\mathrm{\int}^1_{\!\!\! 0} \! \mathrm{d} x  \!\mathrm{\int}^1_{\!\!\! 0} \! \mathrm{d} y  \!\mathrm{\int}^1_{\!\!\! 0} \! \mathrm{d} z    \! \frac{(1\!-\!z)}{z}\! \log \Delta  + C    \, ,
\end{eqnarray}
with $\Delta\!=\![(y(1\!-\!z)q \!+\! z p_3)^2 \!-\!(yq^2\!-\!m^2)(1\!-\!z)\!-\!p_3^2 z]x(1\!-\!x) + m^2 z$. Considering the renormalization conditions that the corrections should be zero at $q^2=4m^2$, the result can be written as
\begin{eqnarray}
\mathcal{T}_\mathrm{P} \! &=& \! \! \frac{(- \mathrm{i}\lambda)^3}{2 (4\pi)^4} \! \!\mathrm{\int}^1_{\!\!\! 0} \! \mathrm{d} x  \!\mathrm{\int}^1_{\!\!\! 0} \! \mathrm{d} y  \!\mathrm{\int}^1_{\!\!\! 0} \! \mathrm{d} z  \frac{1}{z } \Big[(1\!-\!z)\log \Delta \\
&& -\log [(y^2 q^2  \!-\!yq^2\!+\!m^2)x(1\!-\!x)]   \Big]\! - \! C_0  \nonumber \\
 \! &=& \! \! \frac{(- \mathrm{i}\lambda)^3}{2 (4\pi)^4} \! \!\mathrm{\int}^1_{\!\!\! 0} \! \mathrm{d} x  \!\mathrm{\int}^1_{\!\!\! 0} \! \mathrm{d} y  \!\mathrm{\int}^1_{\!\!\! 0} \! \mathrm{d} z  \frac{1}{z } \Big[(1\!-\!z) \! \log \frac{\Delta}{\Delta_0}  \nonumber \\
&& -\log \frac{y^2 q^2  \!-\!yq^2\!+\!m^2 }{(4y^2    \!-\!4y  \!+\!1) m^2}   \Big]    \, , \nonumber
\end{eqnarray}
with $\Delta_0 = [y(1\!-\!z)(y+z-yz)4m^2 \!+\! (z^2 \!-\!z )p_3^2 \!-\!(4y\!-\!1)m^2(1\!-\!z)]x(1\!-\!x)\!+\!m^2z$.

\section{The transition from the traditional method to the UV-free scheme}  \label{int-ex}

For loops with logarithmic divergences, the traditional method of regularization \& renormalization is successful in figuring out the finite loop results. Here is a brief interpretation about the transition from dimensional regularization to the UV-free scheme. Let us set out from an integral with logarithmic divergence in 4 dimensions,
\begin{eqnarray}
I_{4}  = \mathrm{\int}_{\!\!\!\Lambda}  \frac{\mathrm{d}^4k}{(2\pi)^4} \frac{1}{(k^2 \!-\!\Delta )^2}     \, ,
\end{eqnarray}
where $\Lambda$ is a cutoff. The UV-divergent part can be extracted in the dimensional regularization. Taking $\mathcal{T}_\mathrm{F}^d =1/(k^2 \!-\!\Delta )^2$, we will do some operations on the integrand $\mathcal{T}_\mathrm{F}^d $ and see how the above integral turns out in the dimensional regularization, that is,
\begin{eqnarray} \label{drr}
I_D \! &=& \!   \mathrm{\int} \! \frac{\mathrm{d}^Dk}{(2\pi)^D} \Big ( \Big[ \mathrm{\int} \mathrm{d}\xi \frac{\partial  \mathcal{T}_\mathrm{F}^d (\xi)}{\partial \xi} \Big]_{\xi \to 0}  +C \Big) \\
\! &=& \!   \mathrm{\int} \! \frac{\mathrm{d}^D k}{(2\pi)^D}  \Big ( \Big [ \mathrm{\int} \mathrm{d}\xi \frac{-2}{(k^2 \!-\!\Delta + \xi)^3}  \Big]_{\xi \to 0}  +C  \Big ) \, .   \nonumber
\end{eqnarray}
After evaluating the antiderivative with respect to the parameter $\xi$, the result is
\begin{eqnarray}
I_D \! &=& \!   \mathrm{\int} \! \frac{\mathrm{d}^D k}{(2\pi)^D}  \frac{1}{(k^2 \!-\!\Delta )^2}      \, ,
\end{eqnarray}
and we go back to the familiar form in dimensional regularization with the boundary constant $C$ being zero (like tree level). In this case, to figure out the finite result, the procedure of renormalization is required to cancel out the UV divergence as $D \to$ 4.

Let us consider the order in which the momentum integration and the antiderivative are exchanged, and then Eq. (\ref{drr}) becomes
\begin{eqnarray}
I_D' \! &=& \! \Big [ \mathrm{\int} \mathrm{d}\xi \mathrm{\int} \! \frac{\mathrm{d}^D k}{(2\pi)^D} \frac{\partial  \mathcal{T}_\mathrm{F}^d (\xi)}{\partial \xi} \Big]_{\xi \to 0}  + C \\
\! &=& \!  \Big [ \mathrm{\int} \mathrm{d}\xi \mathrm{\int} \! \frac{\mathrm{d}^D k}{(2\pi)^D} \frac{-2}{(k^2 \!-\!\Delta + \xi)^3}  \Big]_{\xi \to 0}   +C  \, .   \nonumber
\end{eqnarray}
At this point the momentum integration is UV converged as $D \to$ 4, and we can evaluate it at $D =$ 4 directly. In this case, one has
\begin{eqnarray}
I_4' \! &=& \!  \Big [ \mathrm{\int} \mathrm{d}\xi \mathrm{\int} \! \frac{\mathrm{d}^4k}{(2\pi)^4} \frac{-2}{(k^2 \!-\!\Delta + \xi)^3}  \Big]_{\xi \to 0}   +C  \, .
\end{eqnarray}
This is the form described by the UV-free scheme. For UV divergence inputs, the evaluation first is the loop momentum $k$ and then is the antiderivative with respect to $\xi$. This non-commutation is a character in quantum theory, and it is the reason why there is no UV divergence in loop calculations in the UV-free scheme.


\end{document}